\documentclass[aps,prb,preprint,groupedaddress,floatfix]{revtex4}
\usepackage{graphicx,amssymb}
\begin{document}

\title{Impact of Friedel oscillations on vapor-liquid equilibria and supercritical properties in $2D$ and $3D$.} 
\author{Caroline Desgranges, Landon Huber and Jerome Delhommelle}
\affiliation{Department of Chemistry, University of North Dakota, Grand Forks ND 58202}
\date{\today}

\begin{abstract}
We determine the impact of the Friedel oscillations on the phase behavior, critical properties and thermodynamic contours in films ($2D$) and bulk phases ($3D$). Using Expanded Wang-Landau simulations, we calculate the grand-canonical partition function and, in turn, the thermodynamic properties of systems modeled with a linear combination of the Lennard-Jones and Dzugutov potentials, weighted by a parameter $X$ ($0<X<1$). Varying $X$ allows us to control the height of the first Friedel oscillation and to provide a complete characterization of the effect of the metal-like character in the potential on the thermodynamic properties over a wide range of conditions. For $3D$ systems, we are able to show that the critical parameters exhibit a linear dependence on $X$ and that the loci for the thermodynamic state points, for which the system shows the same compressibility factor or enthalpy as an ideal gas, are two straight lines spanning the subcritical and supercritical regions of the phase diagram for all $X$ values. Reducing the dimensionality to $2D$ results in a loss of impact of the Friedel oscillation on the critical properties, as evidenced by the virtually constant critical density across the range of $X$ values. Furthermore, our results establish that the straightness of the two ideality lines is retained in $2D$ and is independent from the height of the first Friedel oscillation in the potential. 

\end{abstract}

\maketitle

\section{Introduction}
In recent years, the existence of remarkable contours in the phase diagram, where fluids exhibit properties similar of ideal gases, has drawn considerable interest~\cite{apfelbaum2009correspondence,apfelbaum2009confirmation,apfelbaum2013regarding,nedostup2013asymptotic} and has emerged as a new way to rationalize the properties of supercritical fluids~\cite{sarkisov2002behavior,brazhkin2011van,brazhkin2011widom,brazhkin2013liquid}. This has also provided the basis for new similarity relations and new ways to determine the critical properties for a wide range of fluids~\cite{apfelbaum2009predictions,apfelbaum2012estimate}. Such contours include the Zeno line~\cite{ben1990estimation,kutney2000zeno,kulinskii2010simple,bulavin2011unified,wei2013isomorphism} for which the fluid has the same compressibility factor as an ideal gas. Recent work on metals~\cite{apfelbaum2009predictions,apfelbaum2012estimate,Leanna} has focused on leveraging the apparent straightness of the Zeno line in the low temperature range to determine the critical properties of metals, which are particularly difficult to determine experimentally and exhibit large variations with estimates for e.g. the critical temperature~\cite{Morel} of $Al$ ranging from $5500~K$ to $9600~K$. Other remarkable contours include the $H$ line, which is the curve of ideal enthalpy. Two other contours, underlying the Zeno and $H$ lines, have also been studied in recent work, namely the $S_0$ line, i.e. the curve of maxima for the isothermal compressibility (or, alternatively, the line where the structure factor at zero wave vector $S_0$), and the $H_{min}$ line, i.e. the curve of minima for the enthalpy. These contours exhibit fascinating properties, as the Zeno and $H$ lines remain straight over a wide range of temperatures (typically several hundred of degrees for Argon~\cite{apfelbaum2013regarding}). The $S_0$ and $H_{min}$ contours can also be accurately modeled by simple power laws of the density for Argon~\cite{apfelbaum2013regarding} as well as for nonpolar and quadrupolar molecules~\cite{Abigail} like $SF_6$ and $CO_2$, paving the way for establishing a correspondence between the supercritical properties of different fluids. However, the shape of the thermodynamic contours is known to be very sensitive to the interaction potential between the fluid particles~\cite{nedostup2013asymptotic,kutney2000zeno,Abigail} and it is currently not known how the emergence of a metal-like character in the effective pair potential, characterized by the onset of the Friedel oscillations, impacts these contours. Furthermore, there is no information, to our knowledge, on the effect of the dimension of the system ($2D$ or $3D$) on these contours. 

The aim of this work is to provide a full picture of the impact of the onset of the metal-like character in the pair potential on the thermodynamics at the liquid-vapor phase boundaries and in the supercritical regime, both for bulk phases ($3D$) and in films ($2D$). More specifically, we model here the onset of this metal-like character through a term that mimics the first Friedel oscillation found in the effective pair potential used for metal interactions~\cite{march2005liquid,roth2000fluid,roth2000solid}. This is achieved by taking as the inter-particle potential a linear combination of the Dzugutov potential, weighted by a parameter $X$ ($0<X<1$), and of the Lennard-Jones potential, weighted by the factor $(1-X)$. We focus here on the effect of the Friedel oscilations on the thermodynamic contours at relatively short range (for distances below 3 particle diameters) and use the same potential form for all fluid densities. By varying $X$, we gradually increase  the magnitude of this oscillation and assess its effect on the grand-canonical partition function of the system both in $2D$ and $3D$ and in turn, on all thermodynamic properties. In particular, we focus on elucidating the impact of the metal-like character on the behavior of fluids along the Zeno line, the $H$ line, the $S_0$ line and the $H_{min}$ line.

The paper is organized as follows. In the next section, we present the pair potentials as well as the simulation method used in this work. In particular, we discuss how the recently developed Expanded Wang-Landau simulations~\cite{PartI,PartII,PartIII,PartIV} are applied to determine the grand-canonical partition function in $2D$ and $3D$ and the loci for the coexistence curve and for the thermodynamic contours in the supercritical region of the phase diagram. We then determine the properties of supercritical fluids both for the bulk and for films and assess the impact of the extent of the metal-like character on the phase diagram in both $2D$ and $3D$, before drawing the main conclusion of this work in the last section.

\section{Simulation method}
\subsection{Formalism}
We determine the fluid properties at coexistence and in the supercritical domain of the phase diagram using the recently developed Expanded Wang-Landau (EWL) simulations~\cite{PartI,PartII,PartIII,PartIV}. EWL simulations are based on a flat histogram sampling approach, known as Wang-Landau sampling~\cite{Wang1,Wang2,Shell,Yan,Camp,WLHMC,Tsvetan,DHMD,KennethI,KennethII,Tsvetan2}. They are carried out in the grand-canonical ensemble, within an expanded ensemble approach~\cite{expanded,Lyubartsev,Paul,Shi,Singh,MV1,MV2,Rane1,Rane2,Mercury,Aaron,Erica,Andrew,jctc2015}. This means that the steps for the insertion/deletion of a full particle are achieved according to a staged process by varying the size of a fractional particle. Therefore, the EWL method samples with the same frequency all possible $(N,l)$ values, where $N$ is the number of particles and $l$ is an integer denoting the current stage, or size, of the fractional particle with $0<l<M-1$, where $M$ is the maximum number of stages.

In the EWL method~\cite{PartI}, we consider a simplified expanded grand-canonical ensemble with the following partition function
\begin{equation}
\Theta_{SEGC}(\mu,V,T)= \sum_{N=0}^\infty \sum_{l=0}^{M-1} Q(N,V,T,l) \exp (\beta \mu N)\\
\label{SEGC}   
\end{equation}
with, for $0<l<M$,
\begin{equation}
Q(N,V,T,l)= {V^{N+1} \over { N! \Lambda^{3(N+1)} }}  \int \exp\left(-\beta U({ {\Gamma}})\right) d{ {\Gamma}} \\
\end{equation}

The Metropolis criterion used in the EWL method to accept a move from an old configuration  ($\Gamma_o,N_o,l_o$) to a new configuration ($\Gamma_n,N_n,l_n$) is 
\begin{equation}
acc(o\to n)=min \left[ {1, {{p_{bias}(\Gamma_n,N_n,l_n)} \over {p_{bias}(\Gamma_o,N_o,l_o)}}} \right]
\end{equation}
with the following choice for the biased distribution $p_{bias}$ that ensures the uniform sampling of all $(N,l)$ values
\begin{equation}
p_{bias}(\Gamma,N,l)={{p(\Gamma,N,l)} \over {p(N,l)}}
\end{equation}
where $p(\Gamma,N,l)$ and $p(N,l)$ are joint Boltzmann distributions for $(\Gamma,N,l))$ and $(N,l))$, respectively.

This provides a direct connexion between the biased distribution and $Q(N,V,T,l)$, since the Metropolis criterion becomes
\begin{equation}
acc(o \to n)=min\left[ 1, { Q(N_o,V,T,l_o) V^{N_n} N_o! \Lambda^{3N_o} \exp\left(- \beta U({ {\Gamma}}_n)\right)  \over {Q(N_n,V,T,l_n) V^{N_o} N_n! \Lambda^{3N_n} \exp\left( - \beta U({ {\Gamma}}_o)\right)}}  \right]
\label{Metro2_1}   
\end{equation}
when $l_o$ and $l_n$ are either both equal to $0$ or strictly larger than $0$. Finally, keeping the results for only the cases where $(N,l=0)$ (i.e. systems of $N$ full, 'regular', particles only) allows us to calculate the grand-canonical partition function $\Theta(\mu,V,T)$ as
\begin{equation}
\Theta(\mu,V,T)= \sum_{N=0}^\infty Q(N,V,T,l=0) \exp (\beta \mu N)\\
\label{muVTfinal}   
\end{equation}

\subsection{Models}
In this work, particles interact through a pair potential that is the combination of the Dzugutov ($DZ$) potential~\cite{dzugutov1993formation,dzugutov1992glass,roth2000fluid} and of the Lennard-Jones ($LJ$) potential. The $DZ$ potential exhibits  a minimum for first nearest neighbors and a maximum between the first and second nearest neighbors. This potential mimics the first Friedel oscillation observed in effective pair potentials used to model metals. It has been extensively studied around the fluid-solid transition due to the fact that this potential favors the formation of quasicrystalline phases~\cite{roth2000solid,gebremichael2005spatially,schopf2012embedded,lu2012exploring,mokshin2009shear, achim2014growth,engel2007self,shi2006atomic,ronceray2013influence} and crystalline $\sigma$-phases~\cite{lee2010discovery}. The $DZ$ potential, however, does not exhibit a liquid phase~\cite{roth2000fluid}, and it has been proposed to combine the $LJ$ potential and the $DZ$ potential to design a pair potential that resembles the effective potential of a metal and leads to the existence of a liquid phase. 

\begin{figure}
\begin{center}
\includegraphics*[width=8cm]{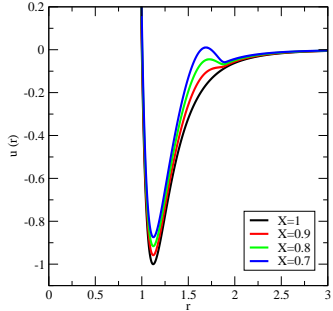}
\end{center}
\caption{Mixed pair potential used in this work for $X=1$, $X=0.9$, $X=0.8$ and $X=0.7$.}
\label{Fig1}
\end{figure}

The resulting pair potential between two particles, separated by a distance $r$, is given by
\begin{equation}
u(r)=X \phi_{LJ}(r)+(1-X) \phi_{DZ}(r)
\end{equation}
where 
\begin{equation}
\phi_{LJ}(r)=4 \left[ {\left({1\over r}\right) ^{12}- \left({1 \over r}\right) ^6} \right]
\label{LJ}   
\end{equation}
and
\begin{equation}
\begin{array}{llll}
\phi_{DZ}(r) & = & \phi_1(r) + \phi_2(r) & \\
\phi_1 (r) & = & A(r^m-B) \exp \left[ {c \over r-a} \right] & r < a\\
 & = & 0 & r>a\\
\phi_2 (r) & = & B \exp \left[ {d \over r-b} \right] & r<b\\
 & = & 0 & r>b\\
 \end{array}
\label{DZ}   
\end{equation}
where $X$ is a weight factor and the potential parameters take the following values $m=16$, $A=5.82$, $C=1.1$, $a=1.87$, $B=1.28$, $d=0.27$ and $b=1.94$. Fig.~\ref{Fig1} shows the impact of the weight factor $X$ on the overall potential energy. 

During the course of the EWL simulations, the interaction between a fractional particle and a full particle is obtained by scaling the parameters of the same dimension as an energy by $(l/M)^{1/3}$  and the parameters of the same dimension as a length by $(l/M)^{1/4}$. We finally add that the same functional forms for the potential are used both in $2D$ and $3D$.

\subsection{Simulation details}
EWL simulations consist of the two types of MC steps, that are attempted with the following rates: 75\% of the attempted MC moves are translations of a single particle (full or fractional) and 25\% of the remaining moves are changes in $(N,l)$ values. For all systems, the maximum number of stages $M$ is set to $100$, the starting value for the convergence factor $f$ in the iterative Wang-Landau scheme is equal to $e$, its final value to $10^{-8}$, with each $(N,l)$ being visited at least 1000 times for a given value of $f$. Simulations are carried out on systems of up to 500 particles for all systems, and the interactions are calculated using a spherical cutoff $(r_c=3)$, with the usual tail corrections applied beyond the cutoff distance~\cite{Allen}. 

\section{Results and Discussion}

\begin{figure}
\begin{center}
\includegraphics*[width=8cm]{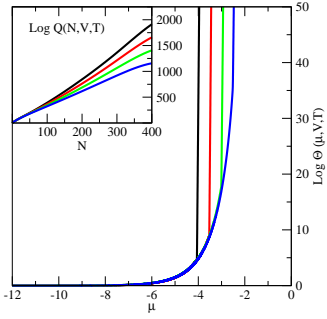}(a)
\includegraphics*[width=8cm]{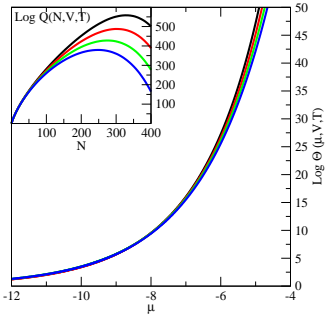}(b)
\end{center}
\caption{Logarithm of the grand-canonical partition function $\Theta(\mu,V,T)$ and of $Q(N,V,T)$ for $3D$ systems: (a) Subcritical fluids ($T=0.85$) and (b) Supercritical fluids ($T=2$). Same legend as in Fig.~\ref{Fig1}.}
\label{Fig2}
\end{figure}

We start by discussing the results on $3D$ systems. The first result we examine is the output from the EWL simulations, i.e. the grand-canonical partition function $\Theta(\mu,V,T)$  and the underlying $Q(N,V,T)$ (see Eq.~\ref{muVTfinal}).  Fig.~\ref{Fig2}(a) shows the grand-canonical partition function for decreasing values of $X$ at a temperature of $T=0.85$. For all systems, $\log \Theta(\mu,V,T)$ exhibits a steep increase associated with the transition from the low density (vapor) to the high density (liquid) phase. The steep increase occurs for increasing value of $\mu$ as $X$ decreases, and the transition on the curve for $\log \Theta(\mu,V,T)$  becomes less and less sharp as $X$ decreases. This behavior can be understood from the plot of $\log Q(N,V,T)$ (inset of  Fig.~\ref{Fig2}(a)) as the partition function $\Theta$ is the sum of  $Q(N,V,T)$ over all possible N values, weighted by the factor $\exp(\beta \mu N)$. The slop of $\log Q(N,V,T)$ is shown to decrease with $X$, and since this slop is equal to $-\mu$, it results in a shift in the liquid to vapor transition towards larger value of $\mu$ as $X$ decreases.  Fig.~\ref{Fig2}(b) shows the behavior of the partition function at a higher temperature ($T=2$). In this plot, the variations of $\log \Theta(\mu,V,T)$ with $\mu$ are dramatically different from those observed at lower temperature. Specifically, $\log \Theta(\mu,V,T)$ does not exhibit a steep increase but rather a steady and smooth increase with $\mu$. The absence of a sharp transition point is characteristic of a supercritical fluid, state which is achieved for all values of $X$ considered here. The behavior for $\log Q(N,V,T)$ is also found to be similar for all $X$ values with a lower maximum observed for $\log Q(N,V,T)$ as $X$ decreases. This results in the slower rate at which $\log \Theta(\mu,V,T)$ increases as a function of $\mu$.

Once the partition function has been determined, the number distribution corresponding to the conditions of coexistence or to the locus of a specific thermodynamic contour, is evaluated as

\begin{equation}
p(N) = {Q(N,V,T)  \exp\left(\beta \mu N\right) \over \Theta(\mu,V,T)}\\
\label{pN}   
\end{equation}

\begin{figure}
\begin{center}
\end{center}
\includegraphics*[width=8cm]{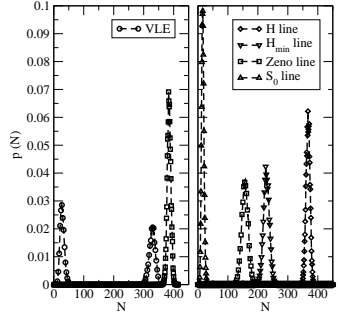}
\caption{Examples of number distribution for $X=1$. (Left) Subcritical fluid ($T=1.1$): vapor and liquid peaks at coexistence and peak for the Zeno line. (Right) Supercritical fluid ($T=2.5$): Peak corresponding to the $S_0$ contour, Zeno line, $H_{min}$ contour and $H$ line.}
\label{Fig3}
\end{figure}

To determine $\mu$ at the vapor-liquid coexistence, we numerically solve the following equation:

\begin{equation}
\sum_{N=0}^{N_b} p(N) = \sum_{N_b}^{\infty} p(N)
\end{equation} 
where $N_b$ is the point at which the function $p(N)$ reaches its minimum, and the left hand side and the right hand side of the equation correspond to the probability of the vapor and of the liquid phase, respectively. The number distribution so obtained at $T=1.1$ is shown for $X=1$ on the left of Fig.~\ref{Fig3}. The loci for the other contours can be determined as follows. For the Zeno line, we numerically solve the following equation:
\begin{equation}
P\bar{V}/RT= \log \Theta(\mu,V,T) /\bar{N}=1
\end{equation} 
where $\bar{V}$ is the reciprocal density and $\bar{N}=\sum N p(N)$ is the average number of particles in the system.

The resulting peak corresponding to the Zeno line at low temperature $T=1.1$ is also shown on the left of Fig.~\ref{Fig3}. The two contours involving the enthalpy can be found by solving two equations for the enthalpy defined as:
\begin{equation}
H=U+PV= {\sum \left(E_{pot}(N)+{3 \over 2}k_B T \right) p(N) \over \sum p(N)}\\+k_BT \log \Theta(\mu,V,T)
\end{equation}
where $E_{pot}(N)$  is the potential energy per particle of a system containing $N$ particles and is collected during the EWL simulation. The $H$ line is then obtained by solving $\bar{H}=5/2 RT$ and the $H_{min}$ contour is obtained by calculating the locus where $H$ reaches its minimum. Finally, the $S_0$ contour is obtained by successive numerical differentiations of $P$ with respect to the number density to achieve $(\partial^2 P/ \partial \rho ^2 )_T=0$. The number distributions obtained at high temperature ($T=2.5$) are shown on the right of Fig.~\ref{Fig3} and exhibit the expected order with the following contours ranging (in the order of increasing $N$ values) $S_0$, Zeno, $H_{min}$ and $H$ lines. We add that some of these contours can only be seen at high (supercritical) temperatures as e.g. the $S_0$ line starts with the critical point and the $H$ line is hidden in the solid domain of the phase diagram at low temperatures.

\begin{figure}
\begin{center}
\includegraphics*[width=8cm]{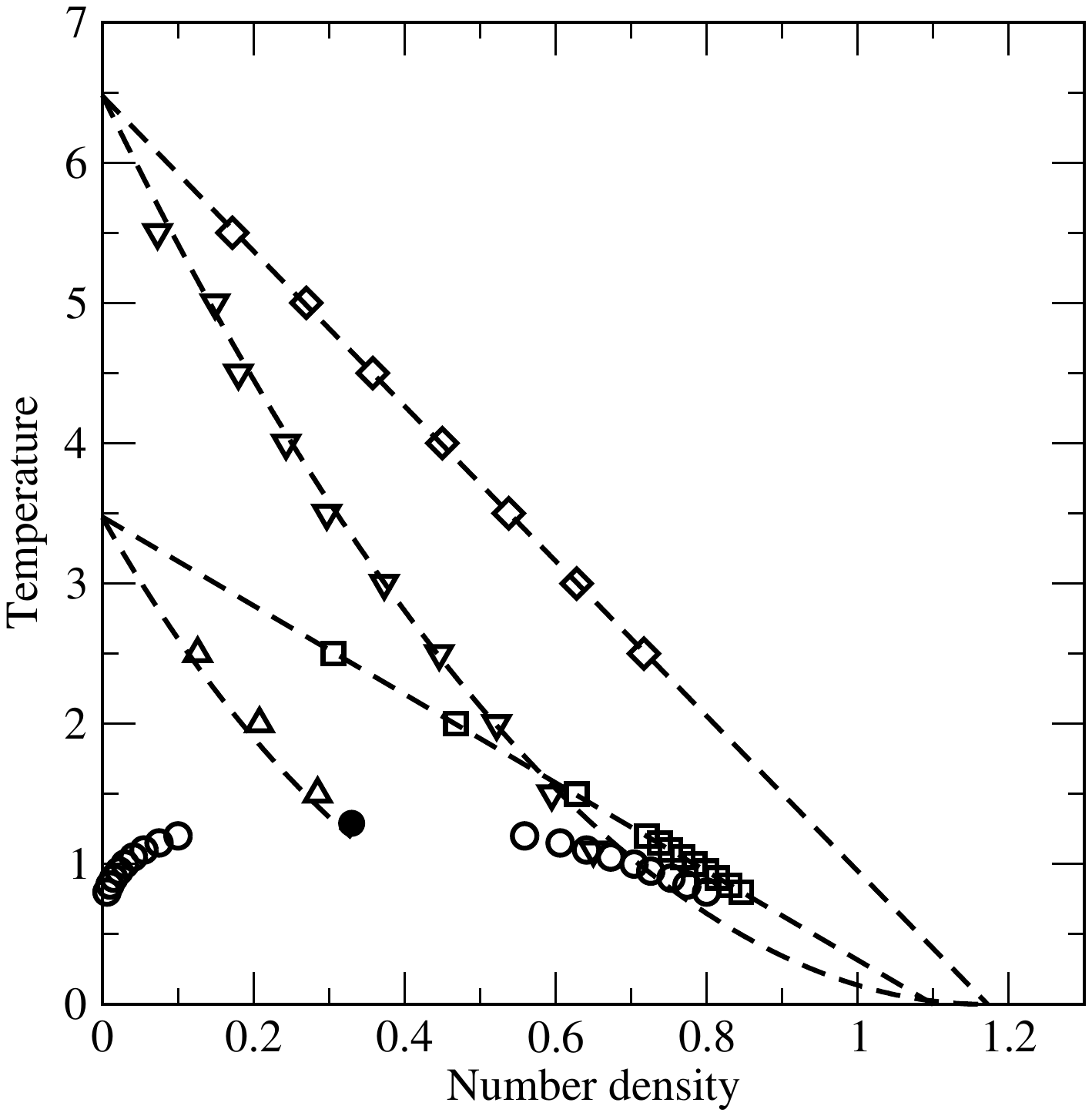}(a)
\includegraphics*[width=8cm]{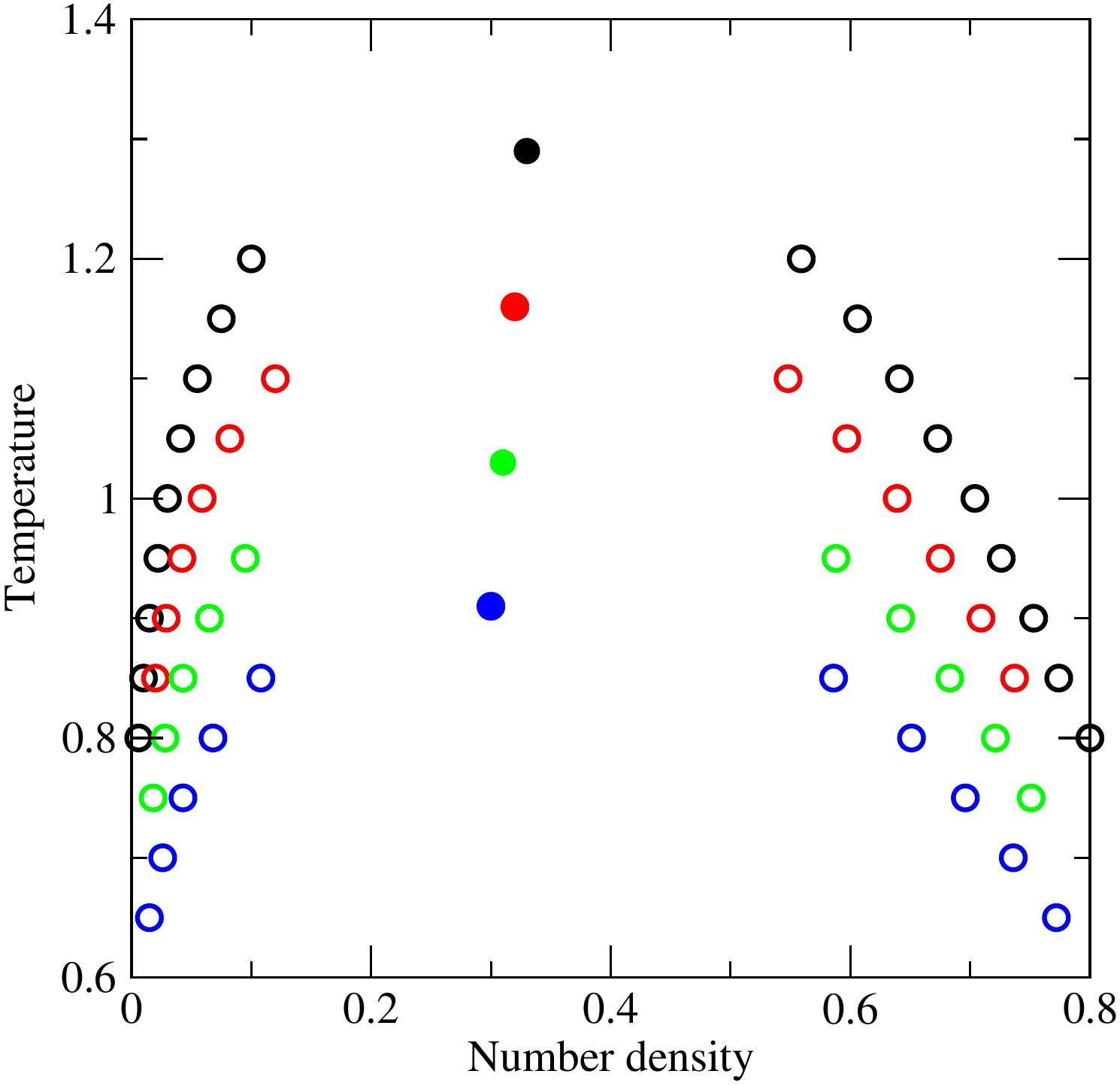}(b)
\end{center}
\caption{(a) Thermodynamic contours in $3D$ for $X=1$ (same legend as in Fig.~\ref{Fig2}), with the critical point shown as a filled circle. The same qualitative behavior is observed for all $X$ values. (b) Coexistence curves and critical points (filled circles) for increasing $X$ values in $3D$ (same legend as in Fig.~\ref{Fig1}). The critical temperature and the critical density exhibit a linear dependence on $X$.}
\label{Fig4}
\end{figure}

The phase diagram and thermodynamic contours for the system $X=1$ are shown in Fig.~\ref{Fig4}(a). Linear regression fits  allow us to determine the Boyle parameters as well as the $H$ parameters (given in Table~\ref{HBC3D}). The Boyle and $H$ parameters we find from EWL simulations are in good agreement with those obtained from density power expansions carried out by Apfelbaum and Vorob'ev~\cite{apfelbaum2013regarding}. Specifically, we find a value for the Boyle temperature $T_B=3.47$ (close to the estimate of $3.42$ from prior work~\cite{apfelbaum2013regarding}) while the EWL Boyle density is of $1.10$ (close to $1.14$ as found previously~\cite{apfelbaum2013regarding}). Similarly we find a $H$ temperature of $6.48$ (slightly above the estimate~\cite{apfelbaum2013regarding} of $6.43$) and a $H$ density of $1.17$ (reasonably close to the value of $1.24$ by Apfelbaum and Vorob'Ev). We also find a behavior for the $S_0$ and $H_{min}$ lines that is consistent with that observed in other work on the van der Waals equation and on Argon. More specifically, we find that the $H_{min}$ line is accurately modeled by the following quadratic law, $T({H_{min}})=T_H(1-\rho(H_{min})/\rho_H)^2$. Similarly, the following cubic law, function of the Boyle parameters, $T({S_0})=T_B(1-\rho(S_0)/\rho_B)^3$ performs very well on the $S_0$ line. We finally determine the critical point from a scaling law for the temperature (with the $3D$ Ising exponent of $0.325$). Our results are in excellent agreement with previous work, with an estimate of 1.29 in this work compared to the estimate of 1.291 using the Transition Matrix Monte Carlo method~\cite{NIST,errington2003direct} or to the estimates obtained through Gibbs Ensemble Monte Carlo simulations~\cite{panagiotopoulos1994molecular,martin1998transferable} of 1.281 and 1.294, respectively. 

The critical density is then obtained from the following similarity law~\cite{apfelbaum2012estimate}:
\begin{equation}
T_c/T_B+ \rho_c/ \rho_B= 0.67
\label{simil}
\end{equation}
We now move on to the impact of the Friedel oscillation on the phase behavior. We show in Fig.~\ref{Fig4}(b) the vapor-liquid coexistence curve for increasing values of $X$. We find that the phase envelope is shifted towards the lower temperatures as $X$ decreases (for $X$ values below $0.6$, no vapor-liquid coexistence can be observed). The presence of the Friedel oscillation notably affects the locus for the phase envelope in terms of temperature, but it also has a significant impact on the symmetry of the coexistence curve (characterized by the different values taken by the two ratios $T_c/T_B$ and $\rho_c/ \rho_B$). More specifically, when using the similarity law of Eq.~\ref{simil}, the $T_c/T_B$ ratio goes from $0.37$ ($X=1$) to $0.38$ ($X=0.9$), $0.40$ ($X=0.8$) and $0.41$ ($X=0.7$). Conversely, the $\rho_c/ \rho_B$ ratio decreases from $0.30$ ($X=1$) to $0.29$ ($X=0.9$), $0.27$ ($X=0.8$) and $0.26$ ($X=0.7$). This change in behavior can be best seen by looking at the variations of the critical temperatures and densities as a function of $X$ which both exhibit an almost perfect linear law. We find that for a given value of $X$, the critical temperature can be modeled as $T_c(X)=1.27 X + 0.018$, while the critical density gives the following linear law $\rho_c(X)=0.1 X + 0.23$. 

\begin{table}[hbpt]
\caption{Boyle, $H$ and critical parameters in $3D$}
\begin{tabular}{|c|c|c|c|c|c|c|c|}
\hline
\hline
$X$ &  $T_B$ &  $\rho_B$ & $T_H$ & $\rho_H$ & $T_c$ & $\rho_c$\\
\hline
\hline
1 & 3.47 & 1.10 & 6.48 & 1.17 & 1.29 & 0.33 \\
0.9 & 3.05 & 1.12 & 5.74 & 1.17 & 1.16 & 0.32 \\
0.8 & 2.58 & 1.15 & 4.94 & 1.18 & 1.03 & 0.31 \\
0.7 & 2.21 & 1.18 & 4.14 & 1.21 & 0.91 & 0.30 \\
\hline
\hline
\end{tabular}
\label{HBC3D}
\end{table}

\begin{figure}
\begin{center}
\includegraphics*[width=8cm]{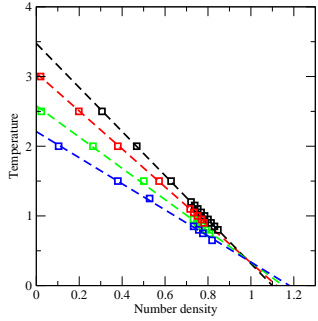}(a)
\includegraphics*[width=8cm]{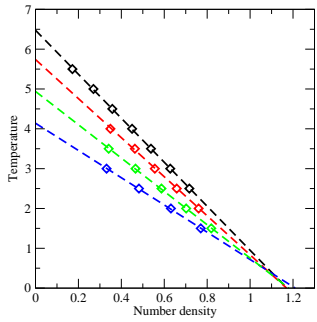}(b)
\end{center}
\caption{Zeno lines (a) and $H$ lines (b) for different $X$ values in $3D$ (same legend as in Fig.~\ref{Fig1}).}
\label{Fig5}
\end{figure}

The Friedel oscillation also impacts the thermodynamic regularity lines as shown for the Zeno line in Fig.~\ref{Fig5}(a) and for the $H$ line in Fig.~\ref{Fig5}(b). Continuing our analysis of the effect of the parameter $X$ on the critical and supercritical properties for the model, we carry out linear fits for the Boyle and $H$ parameters as a function of $X$. We find that the Boyle temperature can be fitted to $T_B(X)=4.25 X - 0.79$, while the Boyle density gives the following linear law $\rho_B(X)=-0.27 X + 1.37$. Similarly, we obtain the following linear fit for the $H$ temperature $T_H(X)=7.82 X - 3.322$ and for the $H$ density $\rho_H(X)=-0.13 X + 1.29$. In line with the trends observed for the vapor-liquid coexistence curve, the increase in the height of the first Friedel oscillation leads to a decrease in the Boyle and $H$ temperatures. However, unlike for the critical densities, we find that decreasing $X$ actually leads to an increase in the Boyle and $H$ densities. This is shown by the crossover point that can be seen on Fig.~\ref{Fig5} for the Boyle and $H$ contours. This means that to achieve ideal gas-like properties, a low temperature metallic system needs to be at a higher density than the corresponding non-metallic system.

\begin{figure}
\begin{center}
\includegraphics*[width=8cm]{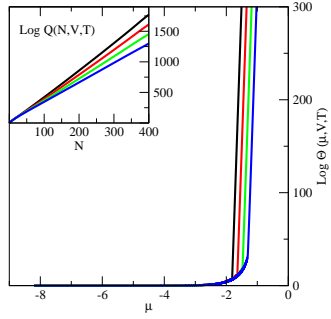}(a)
\includegraphics*[width=8cm]{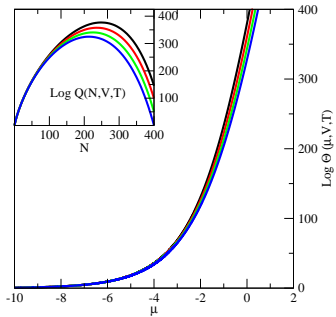}(b)
\end{center}
\caption{Logarithm of the partition function $\Theta(\mu,V,T)$ and of the $Q(N,V,T)$ functions for $2D$ systems: (a) Subcritical fluids ($T=0.41$) and (b) Supercritical fluids ($T=1.5$). Same legend as in Fig.~\ref{Fig1}.}
\label{Fig6}
\end{figure}

We now examine the results obtained in $2D$, for films with increasing heights of the first Friedel oscillation in the potential. Starting with the output from the EWL simulations, we observe the following behaviors. At low temperature (Fig.~\ref{Fig6}(a)), the partition function, plotted as a function of the chemical potential, exhibits a steep increase corresponding to vapor-liquid transition. In line with $3D$ systems, the transition point is shifted towards the larger values for the chemical potential as a result of the increase in the height of the first Friedel oscillation (see e.g. the results shown for $T=0.41$ in Fig.~\ref{Fig6}(a)). This directly stems from the order in which the slopes obtained for $\log Q(N,V,T)$ are obtained for decreasing values of $X$. As for $3D$ systems, this slope is directly related to $-\mu$, and the magnitude of the slopes for $\log Q(N,V,T)$ ($X=1>...>X=0.7$) leads to the order found for the transition points ($X=1$ before $X=0.9$, $X=0.8$ and finally $X=0.7$). At high temperature, we observe a behavior consistent with that found for supercritical $3D$ systems, with the absence of a sharp transition in $\log \Theta(\mu,V,T)$ and the presence of a maximum in $\log Q(N,V,T)$ as a function of $N$. We also find that the maximum reached by $\log Q(N,V,T)$ as a function of $N$ decreases and is reached for lower values of $N$ as $X$ decreases, leading to an earlier increase (in terms of $\mu$) in $\log \Theta(\mu,V,T)$ for larger values of $X$.

\begin{table}[hbpt]
\caption{Boyle, $H$ and critical parameters in $2D$}
\begin{tabular}{|c|c|c|c|c|c|c|c|}
\hline
\hline
$X$ &  $T_B$ &  $\rho_B$ & $T_H$ & $\rho_H$ & $T_c$ & $\rho_c$\\
\hline
\hline
1 & 1.57 & 1.12 & 3.02 & 1.12 & 0.51 & 0.39 \\
0.9 & 1.42 & 1.17 & 2.85 & 1.11 & 0.47 & 0.40 \\
0.8 & 1.28 & 1.20 & 2.44 & 1.19 & 0.44 & 0.39 \\
0.7 & 1.10 & 1.34 & 2.15 & 1.21 & 0.41 & 0.40 \\
\hline
\hline
\end{tabular}
\label{HBC2D}
\end{table}

\begin{figure}
\begin{center}
\includegraphics*[width=8cm]{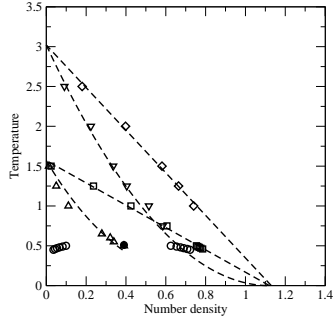}(a)
\includegraphics*[width=8cm]{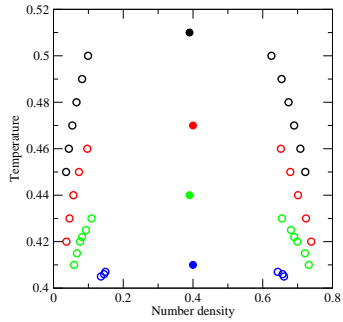}(b)
\end{center}
\caption{(a) Thermodynamic contours for $X=1$ in $2D$ (same legend as in Fig.~\ref{Fig2}), with the critical point shown as a filled circle. The same qualitative behavior is observed for all $X$ values. (b) Coexistence curves and critical points (filled circles) for increasing $X$ values in $2D$ (same legend as in Fig.~\ref{Fig1}). The critical temperature exhibit a linear dependence on $X$, while the critical density remains essentially constant.}
\label{Fig7}
\end{figure}

The phase diagram for  $X=1$ in $2D$, together with the thermodynamic contours, is shown in Fig.~\ref{Fig7}(a). The properties at coexistence as well as the loci for the various contours were determined using the same method as for $3D$ systems, with the exception of the critical temperature that was evaluated through a scaling law with the $2D$ Ising exponent of $1/8$. This scaling law yields a critical temperature ($0.51$) in very good agreement with the estimate of$ 0.515$ obtained in previous work on $2D$ LJ systems by Smit and Frenkel~\cite{Smit}. Overall, we find a much narrower range (than in $3D$) of temperature where the liquid vapor coexistence is observed. This is also, to our knowledge, the first example of the calculation of the Zeno and $H$ line for $2D$ systems. Our results show that the straightness of the Zeno and $H$ lines is indeed conserved in $2D$ as evidenced by the fits presented in Fig.~\ref{Fig7}(a) (with the corresponding Boyle and $H$ parameters given in Table~\ref{HBC2D}). The other two contours ($S_0$ and $H_{min}$ contours) also exhibit a behavior that is consistent with that found for $3D$ systems and are accurately modeled by the simple polynomial laws, function of the Boyle parameters (for $S_0$) and $H$ parameters (for $H_{min}$), as observed for $3D$ systems. Increasing the height of the first Friedel oscillation results in a decrease of the temperature range over which vapor liquid coexistence is observed. This, in turn, results in a steady decrease in the critical temperature (see Fig.~\ref{Fig7}(b)), as evidenced by a linear fit to the EWL data for the critical temperature, which gives the following result $T_c(X)=0.31 X + 0.19$. On the other hand, and, unlike for $3D$ systems, the presence of a first Friedel oscillation in metallic films does not seem to impact the critical density as shown by the virtually constant value of $\rho_c$ obtained for all $2D$ systems.

\begin{figure}
\begin{center}
\includegraphics*[width=8cm]{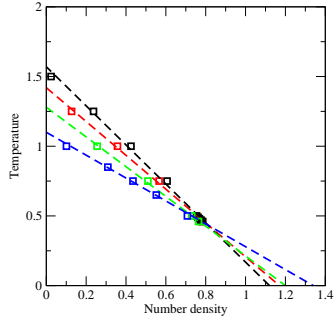}(a)
\includegraphics*[width=8cm]{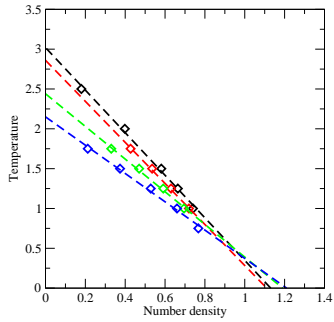}(b)
\end{center}
\caption{Zeno lines (a) and $H$ lines (b) for different $X$ values in $2D$ (same legend as in Fig.~\ref{Fig1}).}
\label{Fig8}
\end{figure}

The Zeno lines and $H$ lines for $2D$ systems are shown in Fig.~\ref{Fig8}(a) and Fig.~\ref{Fig8}(b), respectively. Both sets of contours show that the straightness of these two lines is neither impacted by the reduced dimension of the system ($3D$ to $2D$) nor by the onset of the first Friedel oscillation (with $X$ decreasing from $1$ to $0.7$). We then carry out linear regression fits to the EWL data for the Boyle and $H$ parameters. For the Boyle parameters, we obtain the following linear laws: $T_B(X)=1.55 X + 0.03$ and $\rho_B(X)=-0.69 X + 1.79$. Fig.~\ref{Fig8}(a) shows that the increase in metallic character has qualitatively the same effect as in $3D$ with a steady decrease in the Boyle temperature and an increase in the Boyle density as $X$ becomes smaller. This results in a crossover point for the Zeno lines at a temperature of approximately $0.75$. A similar analysis for the $H$ parameters leads to the following results: $T_H(X)=3.02 X + 0.05$ and $\rho_H(X)=-0.35 X + 1.46$. The decrease in $T_H$ observed in Fig.~\ref{Fig8}(b) is again in line with the results obtained in $3D$, as is the increase in $\rho_H$ as $X$ takes smaller values. Overall, the reduced dimensionality of the system when considering films ($2D$) rather than bulk systems ($3D$) does not dramatically change the thermodynamics of phase coexistence and thermodynamic regularities. However, it mitigates the impact of the Friedel oscillation on the critical and supercritical properties, with much reduced dependence of the critical densities on the $X$ parameter.

\section{Conclusion}
In this work, we carry out EWL simulations to determine the effect of the first Friedel oscillation on the phase behavior, critical properties and thermodynamic regularity contours for films ($2D$) and for bulk phases ($3D$). The onset of the first Friedel oscillation is modeled by superimposing two pair potentials, the Lennard-Jones potential and the Dzugutov potential, weighted by a parameter $X$ ($0<X<1$). The results show that moving away from the LJ system ($X=1$) towards systems with a more pronounced metal-like character ($X<1$) leads to a narrowing of the range of temperature showing vapor-liquid coexistence for both $2D$ and $3D$ systems. For $3D$ systems, increasing the height of the first Friedel oscillation results in a qualitative change in the coexistence curve, with increased asymmetry, as shown by the ratio of the critical parameters to the Boyle parameters, when the metal-like character is increased. This feature is best captured by the almost perfect linear fits exhibited both by the critical temperature and critical density as a function of $X$. Changing the inter-particle potential and making the potential more metal-like does not lead to dramatic changes in the thermodynamic contours, as evidenced by the straightness of the Zeno and $H$ lines that is retained upon decreasing $X$. The results on $2D$ systems are the first, to our knowledge, to examine the behavior of the thermodynamic regularity lines in films and to show that the straightness of the Zeno and $H$ lines is indeed retained upon reducing the dimension of the system. Increasing the height of the first Friedel oscillation in films also leads to the narrowing of the vapor-liquid coexistence curve with decreasing critical temperatures as a consequence of the increase height of the first Friedel oscillation. The loss of a dimension, however, attenuates the impact of the metallic character in the potential as shown by the almost constant critical densities across the range of values of the $X$ parameter. 

{\bf Acknowledgements}
Partial funding for this research was provided by NSF through CAREER award DMR-1052808.\\

\bibliography{ProjectX}

\end{document}